\documentclass[]{jkas} 

\def\year{2023} 
\def\volume{56} 
\def\issue{1} 
\def\beginpage{1} 
\def\received{--} 
\def\accepted{--} 
\def\published{--} 
\date{Received \received; Accepted \accepted; Published \published}

\usepackage{gensymb}
\usepackage{siunitx}
\usepackage{hyperref}
\usepackage{microtype}
\usepackage{xurl}
\usepackage{lineno}
\newcommand{\spherex}{{\textit{SPHEREx}}}
\newcommand{\gaia}{{\textit{Gaia}}}

\title{%
SOLO: wide-field asteroid light curve monitoring system for SPHEREx
}

\author[1,2]{Bumhoo Lim}{0000-0002-8244-4603}
\author[1,2]{Seungwon Choi}{0009-0006-1028-1653}
\author[3]{Yoonsoo P. Bach}{0000-0002-2618-1124}
\author[1,2,$\star$]{Masateru Ishiguro}{0000-0002-7332-2479}
\author[3]{Sunho Jin}{0000-0002-0460-7550}
\author[4]{Carey M. Lisse}{0000-0002-9548-1526}
\author[5]{Max Mahlke}{0000-0003-2831-0513}
\author[6]{Jooyeon Geem}{0000-0002-3291-4056}
\author[1,2]{Jinguk Seo}{}
\author[1,2]{Sihu Ahn}{}
\author[1,2]{Hangbin Jo}{0009-0004-9591-8646}

\affil[1]{Department of Physics and Astronomy, Seoul National University, 1 Gwanak-ro, Gwanak-gu, Seoul 08826, Republic of Korea}
\affil[2]{SNU Astronomy Research Center, Department of Physics and Astronomy, Seoul National University, 1 Gwanak-ro, Gwanak-gu, Seoul 08826, Republic of Korea}
\affil[3]{Korea Astronomy and Space Science Institute, Daejeon 305-348, Republic of Korea}
\affil[4]{Johns Hopkins University Applied Physics Laboratory, 11100 Johns Hopkins Rd, Laurel, MD 20723, USA}
\affil[5]{Universit\'{e} Marie et Louis Pasteur, CNRS, Institut UTINAM (UMR 6213), \'{e}quipe Astro, F-25000 Besan\c{c}on, France}
\affil[6]{Asteroid Engineering Laboratory, Lule\r{a} University of Technology, Kiruna, Sweden}

\def\corrauthor{%
Bumhoo Lim (\email{bumhoo7@snu.ac.kr}) and M. Ishiguro (\email{ishiguro@snu.ac.kr})
}

\def\runningauthor{%
Lim et al.
}

\def\runningtitle{%
Wide-field asteroid light curve monitoring system for SPHEREx
}

\def\keywords{%
Instrumentation: photometers ---
Surveys ---
Minor planets, asteroids: general ---
Techniques: photometric
}

\def\abstracttext{%
We present the Solar system Objects Light curve Observatory (SOLO), a wide-field, high-cadence optical survey system designed to obtain absolutely calibrated asteroid light curves, converted to the {\gaia} $G$-band photometric system, in support of the {\spherex} Solar System Object Catalog (SSOC). SOLO was installed at the Sierra Remote Observatories (SRO) in California, USA, in July 2025 and is optimized for continuous, multi-night monitoring of asteroid brightness variations. We describe the system configuration, remote operation, and data reduction pipeline, and evaluate its optical and photometric performance using commissioning data. SOLO achieves stable photometric calibration across the $11.6 \deg^2$ field of view and reaches a 10-$\sigma$ limiting magnitude of $G \sim 17.5$ for a 180~sec exposure. Sample asteroid light curves obtained over multiple nights demonstrate consistent absolute photometry at the same rotational phase, validating the estimated performance. Finally, we outline the planned operational use of SOLO in connection with NASA's {\spherex} mission. Full science operations of SOLO are scheduled to begin in January 2026. Using these data, we aim to obtain on the order of $10^{3}$ absolutely calibrated asteroid light curves per year in the {\gaia} $G$-band, which will be used to support the construction and scientific utilization of the {\spherex} SSOC.
}

\begin{document}
\jkashead 



\section{Introduction}\label{sec:intro}

Small bodies in the Solar System (e.g., asteroids and comet nuclei) are primitive remnants of its formation. Most of them are smaller than a few tens of kilometers in diameter and therefore typically exhibit irregular shapes that cause periodic brightness variations due to their rotation or tumbling motion (the so-called light curves) and due to changes in the solar phase angles (i.e., the angle between Sun---target---observer). Light curves reveal key physical properties of asteroids, such as shape, rotation period ($P_\mathrm{rot}$), and spin-axis orientation. Since their rotational states reflect collisional evolution and thermal effects such as the YORP torque, investigating their light curves is of great importance for understanding the physical and dynamical evolution of small bodies in the Solar System (e.g., \citealt{1985Icar...62...30D, 2008Icar..197..497P}).

To obtain reliable light curves, it is important to measure asteroid fluxes at short time intervals of a few minutes, given that typical rotation periods range from several hours to several tens of hours \citep{2000Icar..148...12P, 2009Icar..202..134W}. Except for a small number of fast-rotating asteroids with periods below one hour, such temporal resolution is sufficient for resolving rotational brightness variations. Conventionally, light curve observations have been performed for a limited number of targets of particular interest using "relative" photometry with comparison stars in the same field. These relative photometric data have been compiled in the Database of Asteroid Models from Inversion Techniques (DAMIT; \citealt{2018A&A...617A..57D}) and the Asteroid Lightcurve Database (LCDB; \citealt{2009Icar..202..134W}), which have been used to reconstruct three-dimensional asteroid shape models through light curve inversion techniques \citep{2010A&A...513A..46D}. Even this most comprehensive catalog (i.e., DAMIT) contains, as of January 2026, well-established light curves for only $\sim300$ asteroids (defined by the \texttt{quality flag}\footnote{\url{https://damit.cuni.cz/projects/damit/pages/documentation}} $\geq2.5$), representing only $\sim0.01~\%$ of the known population. This clearly demonstrates that the number of asteroids with reliably determined light curves remains very limited.

In recent years, large-scale sky surveys such as the Pan-STARRS \citep{2016arXiv161205560C}, the Zwicky Transient Facility (ZTF; \citealt{2019PASP..131a8002B},  and the Vera C. Rubin Observatory Legacy Survey of Space and Time (LSST; \citealt{2019ApJ...873..111I}) have made it possible to provide time-series photometric data of hundreds of thousands of asteroids with excellent sensitivity. However, their cadence is generally limited (on the order of a few observations per day), which is insufficient to construct accurate light curves. Nevertheless, steady efforts have been made to estimate spin-axis orientations and rotation periods by simultaneously modeling the three-dimensional shape and the phase function (see, e.g., \citealt{2024A&A...687A..38C}).

The rotational properties inferred from asteroid light curves are essential to
obtain further physical properties. In 2024, our research group launched a program to construct reflectance spectra of small Solar System bodies using data from the Spectro-Photometer for the History of the Universe, Epoch of Reionization, and Ices Explorer (\spherex) mission \citep{2020SPIE11443E..0IC, 2022Icar..37114696I,2025arXiv251102985B}. Launched in March 2025, {\spherex} will carry out a long-term spectroscopic survey for 2 years covering the wavelength range of 0.75–5.0 $\micro$m, which will make it possible to produce spectral data of individual objects by measuring their fluxes at different wavelengths and at different epochs during repeated all-sky scans. However, to utilize these {\spherex} data for studies of asteroids, it is essential to characterize their rotational brightness modulation and changes in phase angle. This is because, during the {\spherex} spectral scans, asteroids rotate multiple times and the phase angle changes, producing flux variation that cannot be readily distinguished from intrinsic spectral features.

{\spherex} will observe more than $\approx10^5$ small Solar System bodies \citep{2020SPIE11443E..0IC}, which is much larger than the number of asteroids with observed light curves. To enable detailed spectral characterization via light curve analysis, we constructed a new telescope system, the Solar system Objects Light curve Observatory (SOLO), at the Sierra Remote Observatories (SRO) in California, USA, in July 2025. Equipped with a wide field-of-view (FoV) of approximately $11.6~\mathrm{deg}^2$, SOLO is dedicated to monitoring asteroid light curves with high cadence. In particular, over the next two years, SOLO aims to provide ancillary light curve data for the {\spherex} Solar System Objects Catalog ({\spherex} SSOC). This will be achieved by coordinating nearly simultaneous ground-based observations with the {\spherex} survey regions, enabling us to trace and correct the flux variability of asteroids within the {\spherex} field of view and thereby improve the reliability of the corrected {\spherex} spectral database (Section~\ref{sec:science_programs}).

This paper presents an overview of the SOLO system. Section~\ref{sec:characteristics} summarizes the system's configuration and technical characteristics. Section~\ref{sec:performance} describes the system performance, the data calibration pipeline, and presents example asteroid light curves obtained with SOLO. Section~\ref{sec:science_programs} outlines our observing strategy to maximize light curve acquisition and highlights SOLO's contribution to correcting {\spherex} data.

\section{System Characteristics}\label{sec:characteristics}

In this section, we summarize the system characteristics of SOLO. Section~\ref{subsec:telescope_and_detector} provides the specifications of the telescope, detector, and filter, and the details of the observatory site. Section~\ref{subsec:operation} describes the operational system and data reduction pipeline.

\subsection{Site and Instrument Overview}
\label{subsec:telescope_and_detector}

\begin{figure}
\centering
\includegraphics[width=85mm]{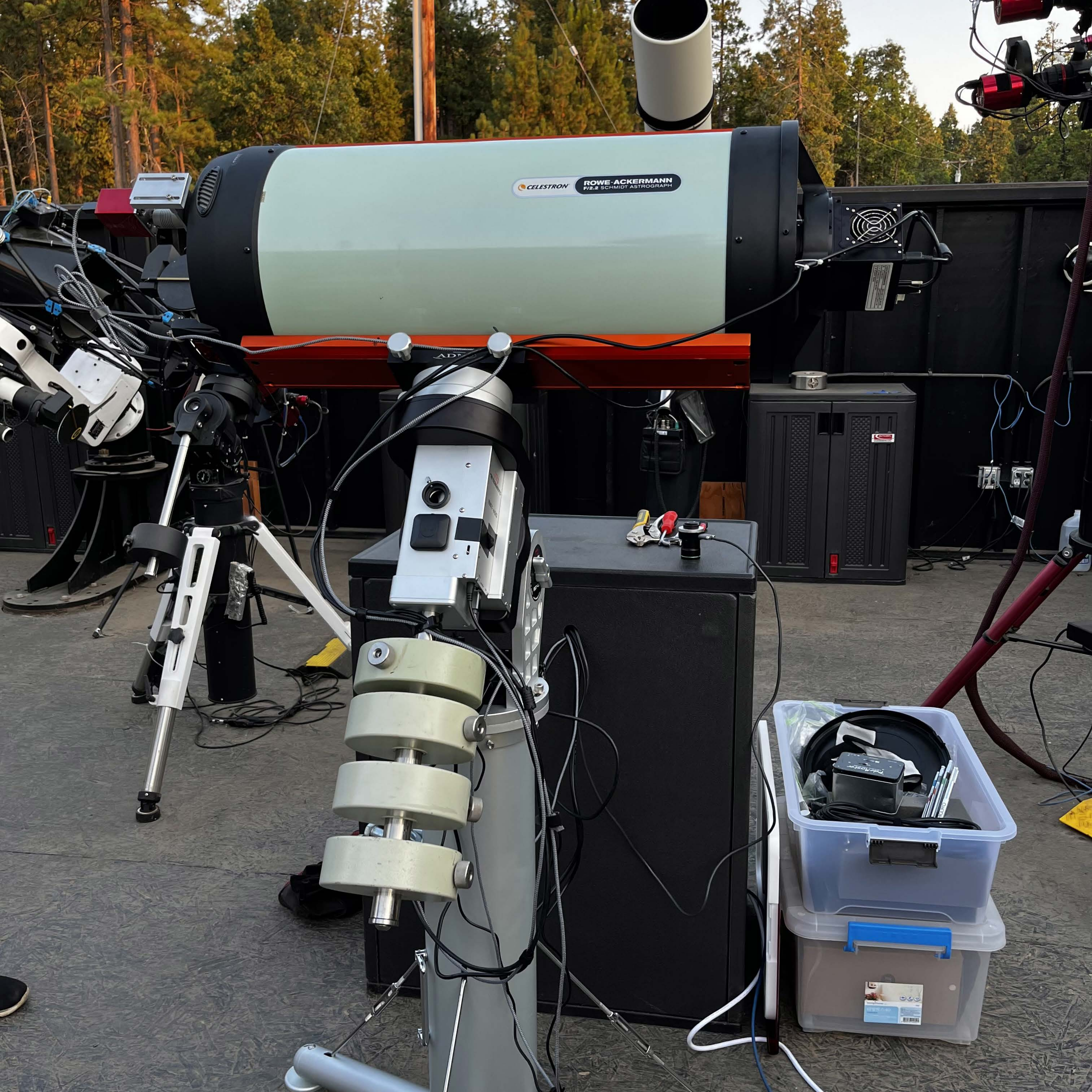}
\caption{Overview of the SOLO observation system after its installation at the Sierra Remote Observatories (SRO) in California, USA, in July 2025.\label{fig:solo}}
\end{figure}

\begin{table}[t!]
\caption{Specifications of the SOLO telescope (RASA-11) and CMOS camera (Kepler 4040FI) \label{tab:solo_spec}}
\centering
\begin{tabular}{lr}
\midrule
Effective diameter & 279 mm (11-inch) \\ 
Focal length & 620 mm \\ 
f-ratio & f/2.22 \\ \addlinespace
Pixel array & 4096 $\times$ 4096 \\
Pixel size & 9 $\mathrm{\mu m}$ \\
Pixel scale & $2^{\prime\prime}.97$ $ \mathrm{pixel^{-1}}$ \\ 
FoV & 3\degree.4 $\times$ 3\degree.4 \ (11.6 $\deg^2$)\\ 
Readout time & $\sim2$ s \\
Readout noise & 34.1 $\mathrm{e^{-}}$ \\
Full well & $4096$ ADU \\
Linearity & $\sim3800$ ADU \\
Effective gain & 18.69 $\mathrm{e^{-}/ADU}$ \\
Dark current$^{\rm a}$ & 0.08 $\mathrm{e^{-}\ pixel^{-1} s^{-1}}$ \\ 
\bottomrule
\end{tabular}
\tabnote{
The camera specifications (readout noise, full well, linearity, effective gain, and dark current) are the values at $1\times1$ HDR mode and $2.8\times$ PGA gain setting.
\\ $^{\rm a}$ The dark current was measured at $-10\degree$C.
}
\end{table}

\begin{figure}
\centering
\includegraphics[width=85mm]{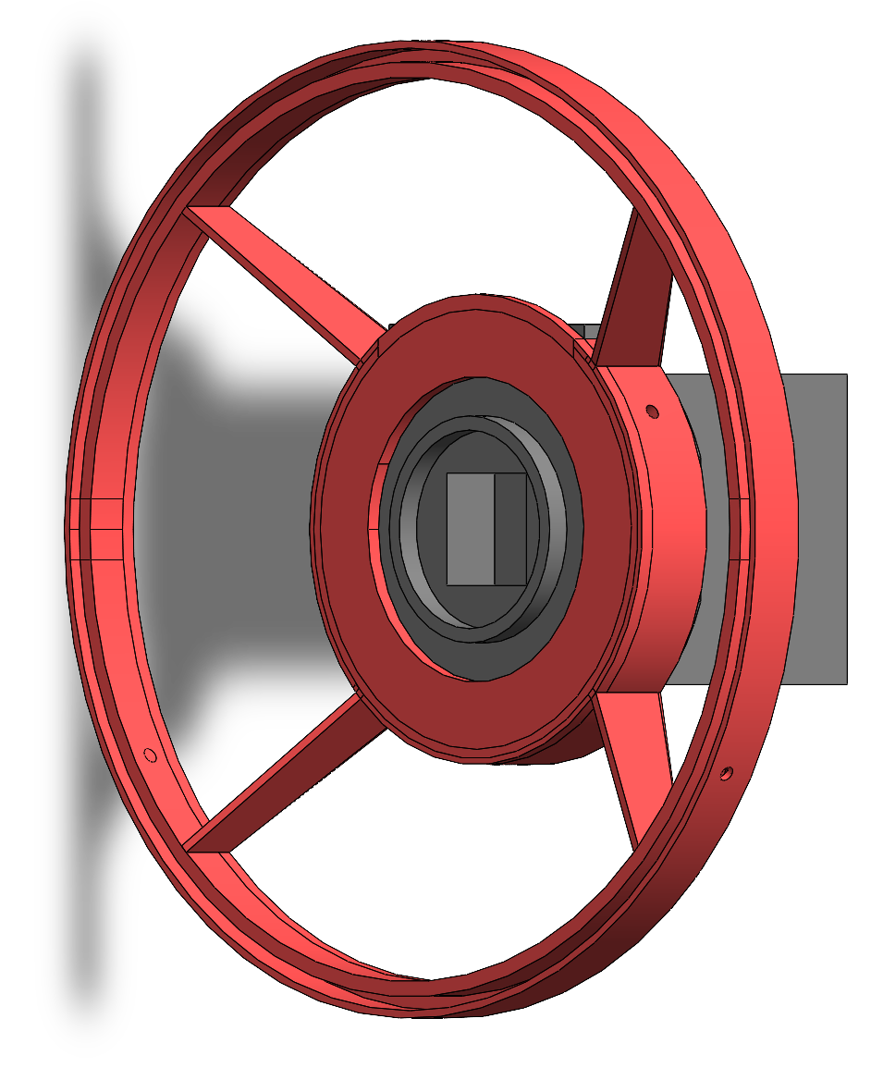}
\caption{Design of the custom spider assembly for SOLO (shown in red). The circular interface on the front-left side attaches to the telescope entrance aperture, while the dark-gray structure on the rear-right side represents the mounted CMOS camera. A faint outline of the spider is shown in the background to illustrate its projected side view and overall geometry. \label{fig:spider}}
\end{figure}

\begin{figure}
\centering
\includegraphics[width=85mm]{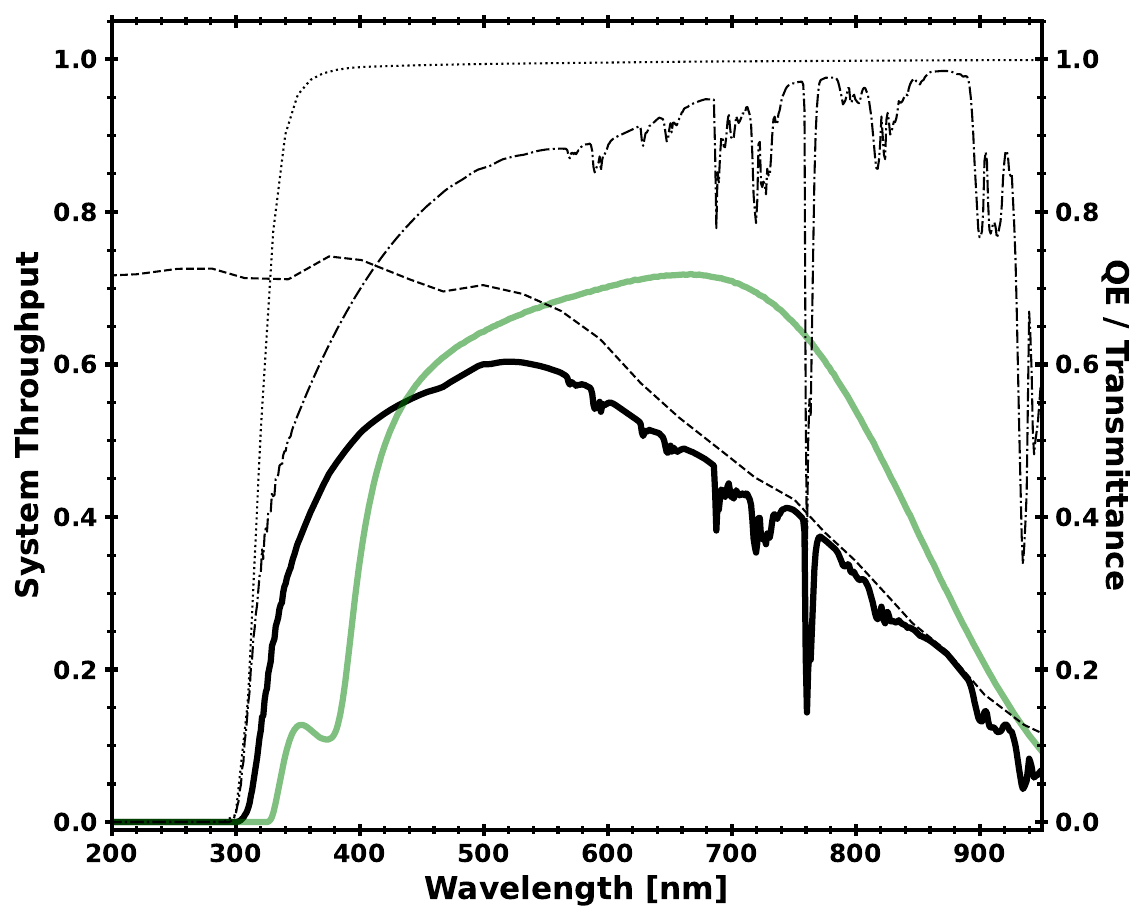}
\caption{Expected system throughput of SOLO, considering detector QE (dashed), filter transmittance (dotted), and atmospheric transmission (dash-dotted). 
The black solid line shows their product (i.e., system throughput). For comparison, the green line shows the filter transmittance of {\gaia} $G$-band. We adapted the atmospheric model constructed by \texttt{lowtran} \citep{1988ugls.rept.....K}. \label{fig:throughput}}
\end{figure}

Figure \ref{fig:solo} shows the SOLO system after its installation. Considering the short development period of about one year, the observing system was designed using a combination of commercially available components. Based on the compatibility between available commercial telescopes and instruments, as well as the FoV matching with single detector of {\spherex} (3\degree.5 × 3\degree.5; \citealt{2025arXiv251102985B}), we adopted an 11-inch Rowe-Ackermann Schmidt Astrograph (RASA-11) manufactured by Celestron\footnote{\url{https://www.celestron.com/}}. The telescope has a prime-focus optical system with an effective diameter of 0.28 meters (11 inches). With this design, the telescope achieves fast optics with an effective focal ratio of f/2.2, providing good image quality over a 3-degree diameter circle without notable distortion. Combined with ancillary corrective lenses, it delivers a coma-free flat field without significant optical aberration over a 43 mm diameter circle on the focal plane. The telescope is paired with a RainbowAstro\footnote{\url{https://www.rainbowastro.com/}} RST-400 mount. Because the fast optics provide relatively low spatial resolution and the mount offers sufficient tracking performance within the given exposure times (see below), we have not implemented an auto-guiding system at this stage in order to maintain simplicity of operation.

We adopted the Kepler 4040 FI camera manufactured by Finger Lakes Instrumentation\footnote{\url{https://www.flicamera.com/}} (FLI) as the focal-plane detector, which provides a large-format sensor while being relatively easy to obtain. The camera employs a GSense4040 front-illuminated CMOS sensor with a peak quantum efficiency (QE) of 70 \%, featuring a $4096 \times 4096$ pixel array with 9~\micro m pixels. This configuration yields a pixel scale of $2^{\prime\prime}.97$ and provides a wide FoV of 3\degree.4 × 3\degree.4. The key system specifications are summarized in Table \ref{tab:solo_spec}.

It should be noted that when this large-format camera is coupled with this fast optical system, mechanical flexure of the assembly can cause focal-plane tilt or small variations in focus depending on the telescope orientation. To ensure mechanical stability, we developed a custom-designed camera holder (or “spider”, Fig.~\ref{fig:spider}) that rigidly supports the CMOS camera at the entrance aperture and maintains its parallelism with the focal plane.
Finite-element calculations indicate that flexure in the X–Y plane corresponding to field shifts on the detector is more than two orders of magnitude smaller than the pixel size and, therefore, negligible. Potential flexure along the Z direction (i.e., focus shift) was assessed experimentally. As shown in Section~\ref{subsec:optical_performance}, the point spread function (PSF) remains stable across the FoV, demonstrating that any focus variation induced by the camera load is negligible in practice. These facts confirm that the spider assembly introduces no measurable degradation in either PSF uniformity or photometric performance.

Since the primary goal of SOLO is to correct the apparent brightness variations of asteroids caused by their rotation (i.e., changes in projected cross section), we adopted a single-filter system to achieve high photometric precision and maximize observing efficiency. We installed a SCHOTT WG320 long-pass filter manufactured by Edmund Optics\footnote{\url{https://www.edmundoptics.co.kr}}. This UV-cut clear filter provides high optical throughput across the visible range, which is well suited for maximizing the photometric sensitivity of SOLO. The combined system throughput (taking into account the detector quantum efficiency, filter transmittance, and the typical atmospheric transmission) is shown in Figure~\ref{fig:throughput}. Although the detailed efficiency curves for the RASA-11 corrector plate and primary mirror are not publicly available, the manufacture (Celestron) indicates that the telescope efficiency decreases toward wavelengths below 400 nm. Overall, as a result, the effective bandpass of SOLO is designed to roughly match that of the {\gaia} $G$-band (green line in Figure~\ref{fig:throughput}), and this consistency is quantitatively demonstrated in Section \ref{subsec:calibration} through an absolute photometric calibration between SOLO and the {\gaia} $G$ magnitudes at the $\approx3 \%$ level.

The system is hosted at SRO in the Sierra Nevada Mountains, California, USA (37\degree.07N, 119\degree.40W; 1405 m altitude). According to information provided by SRO\footnote{\url{https://www.sierra-remote.com/}}, the site offers excellent observing conditions, with approximately 290 clear nights annually and a measured sky brightness of 21.78 $\rm{mag\ arcsec^{-2}}$ in the $V$-band. The typical seeing is $\sim 1.5^{\prime\prime}$. The selection of the SRO site was primarily motivated by its high clear-sky fraction and dark sky background. Because the primary goal of SOLO is to obtain time-series photometry in coordination with the {\spherex} observations over as many nights as possible, a site with excellent weather statistics was essential. In addition, our detector (see below) has a relatively large pixel scale, making the signal-to-noise ratio (S/N) of faint targets highly sensitive to the night-sky brightness. Considering these requirements, we evaluated several candidate sites both within and outside South Korea, and finally selected SRO as the site that best satisfies these conditions. SRO provides roll-off-roof dome infrastructure and on-site maintenance as part of its telescope-hosting service and, we operate SOLO alongside other remotely managed telescopes within this facility. 

\subsection{System operation and data reduction}\label{subsec:operation}

\begin{figure*}
\centering
\includegraphics[width=\linewidth, trim={0cm 5cm 0cm 6cm}, clip]{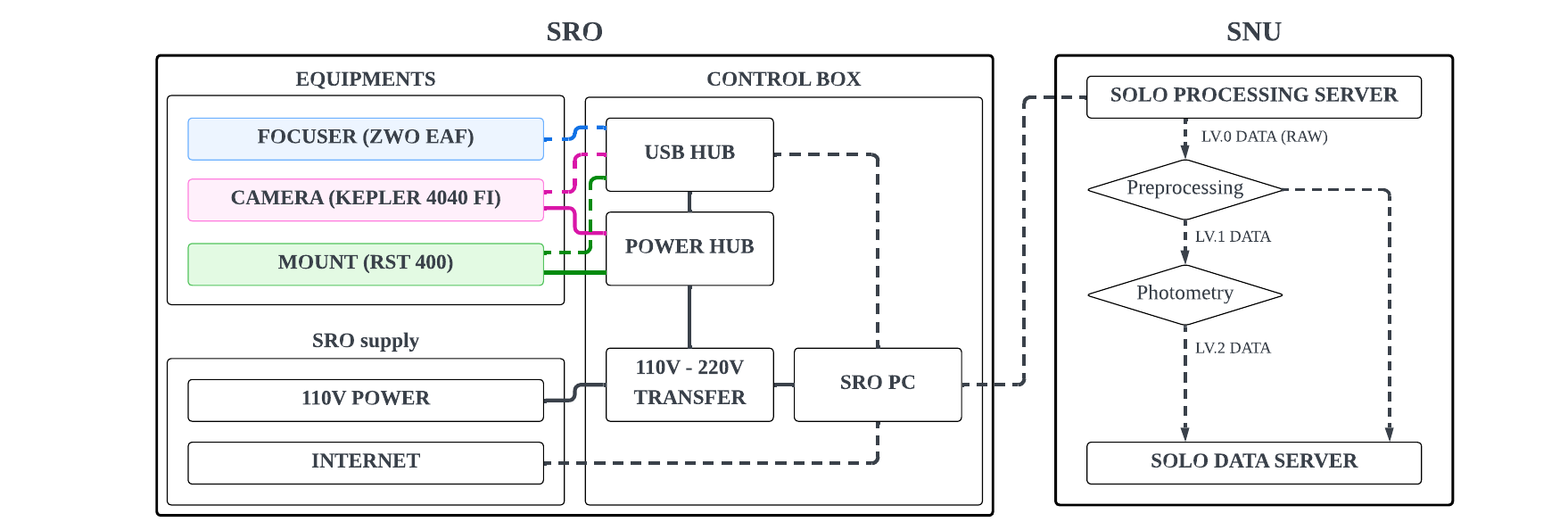}
\caption{Schematic diagram of the SOLO system configuration and data-processing pipeline. Solid and dashed lines indicate power and data connections, respectively. The three observational subsystems (Equipments), the RST-400 mount, the Kepler 4040 FI camera, and the ZWO EAF focuser, are installed at Sierra Remote Observatories (SRO) and are controlled by an on-site PC located in the control box. The SRO PC is dedicated to device control and data acquisition and receives power supply and internet connectivity from the SRO infrastructure. The system is remotely operated from Seoul National University (SNU), South Korea, via the internet. The raw data acquired at SRO are transferred to a dedicated SOLO processing server at SNU, where the primary data reduction and photometric analysis are performed, and the resulting data products are stored in the SOLO data server.
 \label{fig:tcs}}
\end{figure*}

\begin{figure*}
\centering
\includegraphics[width=\linewidth]{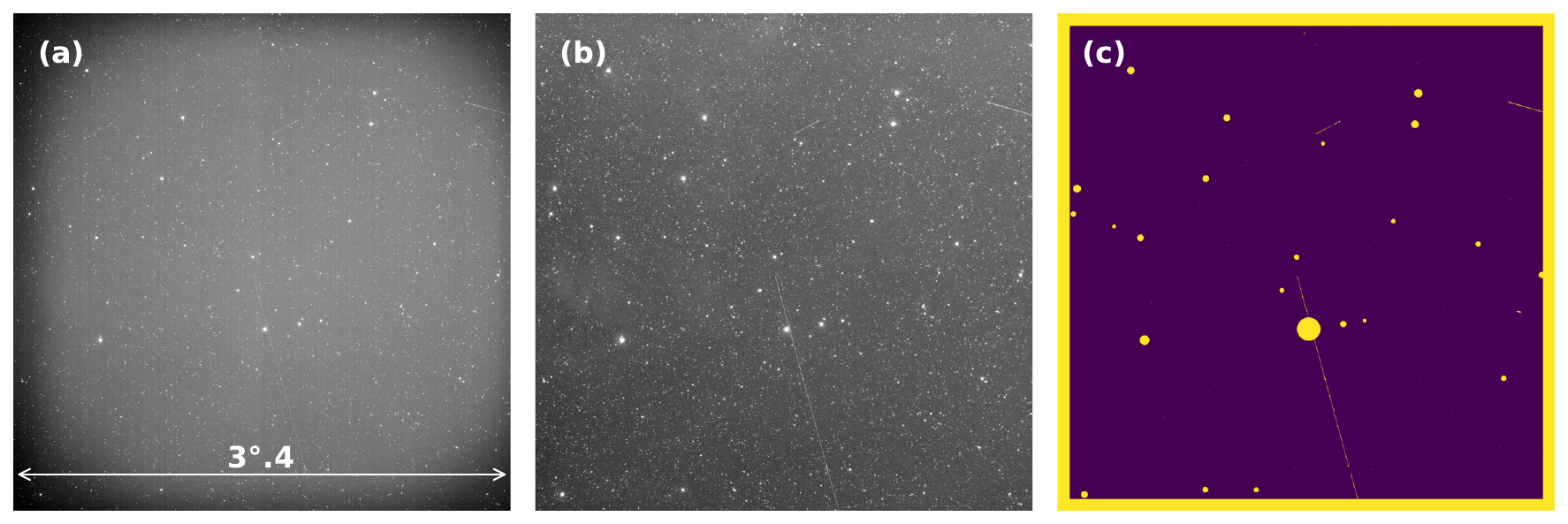}
\caption{An example image from SOLO, taken on July 22, 2025, with 180~sec exposure. (a) Raw image, (b) calibrated image, and (c) mask image. In panel (a), strong vignetting is visible in the four corners, along with additional shading on the top, bottom, and left edges caused by the square filter holder. After applying the flat‐field correction, these effects are largely removed in panel (b), resulting in a much more uniform background across the entire field of view. Panel (c) shows an example of the mask image, 
where bright stars, satellite trails, and bad pixels are masked out to exclude them as outliers during the data calibration.\label{fig:lv0lv1}}
\end{figure*}

Overall operation and data process is shown in Figure \ref{fig:tcs}. SOLO is operated remotely via a custom telescope control system (TCS). Located at SRO, the system is managed through remote access to an on-site host computer ("Control Box" in Figure \ref{fig:tcs}). The three instruments, comprising the mount (RST-400), the camera (Kepler 4040 FI), and ZWO\footnote{\url{https://www.zwoastro.com/}} EAF focuser, are controlled via the Astronomy Common Object Model\footnote{\url{https://ascom-standards.org/}} (ASCOM) Alpaca protocol, which provides a standardized interface for astronomical device communication. Using this protocol, we constructed a Python-based TCS that issues Alpaca commands through a command-line interface. Through this interface, we not only control basic device functions but also execute complex routines such as auto-focusing, field identification, and acquisition of calibration frames (e.g., bias and dark images). This system enables fully robotic operation of the SOLO survey observations (Section~\ref{sec:science_programs}).

At the end of each night, the SOLO data are processed through a dedicated reduction pipeline that performs the standard calibrations required for astronomical CMOS data. To reduce network traffic and computational load at SRO, these data processing tasks are executed on a dedicated SOLO processing server installed at Seoul National University (SNU; see the right panel of Figure \ref{fig:tcs}). The procedure begins with astrometric calibration (WCS solution), followed by bias subtraction, dark correction, and flat-fielding (Figure \ref{fig:lv0lv1}b). We then apply a bad-pixel mask to exclude problematic regions such as saturated pixels, hot pixels (dark current $> 10\ \mathrm{e^{-}\ pixel^{-1}\ s^{-1}}$), and sensor edges (Figure \ref{fig:lv0lv1}c). Here, we also remove the linear features (primarily resulting from satellite trails) and pixels likely affected by nearby bright sources identified using \texttt{sep} \citep{1996A&AS..117..393B, 2016JOSS....1...58B}. Following calibration, we determine the photometric zero point by cross-matching field stars with {\gaia} Data Release 3 (DR3) $G$-band sources \citep{2023A&A...674A...1G}, excluding bad pixels from this estimate (Lv.1 data in Figure \ref{fig:tcs}). Afterward, asteroid candidates in the field are identified, and aperture photometry is performed to derive their instrumental magnitudes. These are then calibrated to apparent magnitudes using the derived zero point (Lv.2 data in Figure \ref{fig:tcs}).

All data reduction processes and the documentation are implemented in the Python-based pipeline \texttt{solopy}\footnote{\url{https://github.com/bumhooLIM/solopy}}, which makes extensive use of widely adopted astronomical libraries such as \texttt{astropy} \citep{2013A&A...558A..33A} and \texttt{astroquery} \citep{2019AJ....157...98G}, along with other supporting packages \citep{1996A&AS..117..393B, 2015ascl.soft10007C, 2016JOSS....1...58B}.

\section{Performance evaluation}\label{sec:performance}

In this section, we report the initial performance analysis of the SOLO system. Given its wide field of view and fast optics, we first investigated its optical performance, including vignetting and the PSF, in Section \ref{subsec:optical_performance}. We then performed the absolute photometric calibration against the \textit{Gaia} sources in Section \ref{subsec:calibration}. Finally, we estimate the system's limiting magnitude (Section \ref{subsec:limitmag}) and present sample asteroid light curves (Section \ref{subsec:sample_lightcurve}).

\subsection{Optical performance}\label{subsec:optical_performance}

\begin{figure}
\centering
\includegraphics[width=\linewidth]{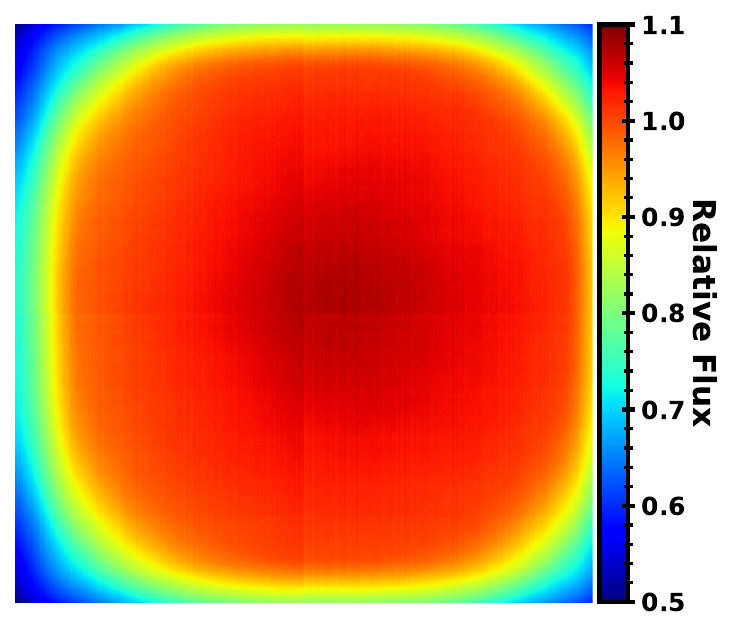}
\caption{Median-stacked and normalized master sky flat of SOLO. Each individual frame was normalized by its respective median value prior to stacking, resulting in a master flat with a median value of approximately unity.
Toward the field edges, the relative throughput decreases to $\lesssim 70-80$ \% of the reference value; however, this spatial sensitivity variation can be properly corrected using this median-stacked flat field, albeit with a corresponding reduction in signal-to-noise ratio.
\label{fig:skyflat}}
\end{figure}

\begin{figure}
\centering
\includegraphics[width=85mm]{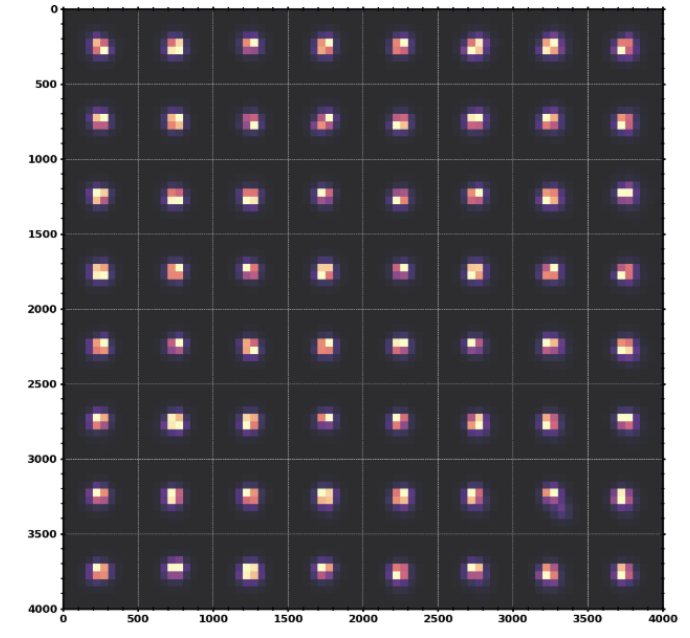}
\caption{Spatial distribution of the point spread function (PSF) across the SOLO field of view. The full field is divided into $500 \times 500$ pixel subregions, with each panel displaying a mean-stacked PSF image for its respective area. Given the large pixel scale of SOLO ($2^{\prime\prime}.97$ pixel$^{-1}$), the majority of the PSF flux is confined within only a few pixels in all subregions, indicating a compact and spatially consistent PSF over the field of view.\label{fig:psf}}
\end{figure}

To evaluate the optical performance of SOLO, we first constructed a master sky flat to characterize the system's vignetting (Figure \ref{fig:skyflat}). 
Because of the wide field of view, standard twilight flats were found to be affected by sky gradients and thus unsuitable for flat-fielding. In such a wide-field system as SOLO, the twilight sky brightness is not spatially uniform across the field, leading to systematic flat-field errors if used directly.
Alternatively, we generated the master flat using 1,132 night-sky images taken between 2025 July 17 and August 30. These frames were selected to have exposure times longer than $60$ seconds and elevations $> 50\degree$ to ensure a uniform background. The master flat was created by median-stacking these data after applying sigma clipping and masking bad pixels and stellar sources (identified using \texttt{sep}; \citealt{1996A&AS..117..393B, 2016JOSS....1...58B}).

The resulting flat field shows a consistent response across most of the FoV. However, significant vignetting is present near the edges, where the relative flux decreases to $<70 \%$ of the central value. This strong vignetting arises because the telescope’s corrected image circle (43.3~mm) is smaller than the sensor’s diagonal (52~mm), which is consistent with the manufacturer’s specifications for the Celestron RASA-11 (Celestron Inc., 2025\footnote{\url{https://www.celestron.com/products/11-rowe-ackermann-schmidt-astrograph-rasa-11-v2-optical-tube-assembly-cge-dovetail}}). It should be noted that the region outside the manufacturer-defined optimized image circle does not represent a hard cutoff for data acquisition. Rather, the throughput gradually decreases toward the field edges, leading to reduced illumination rather than a loss of usable data. In this work, we therefore attempt to utilize photometric data beyond the telescope’s corrected image circle (43.3~mm) by applying an appropriate flat-field correction. While this vignetting pattern is evident in the raw image (Figure \ref{fig:lv0lv1}a), it is largely corrected by the master flat (Figure \ref{fig:lv0lv1}b), rendering the data at the field edges scientifically useful.

The optical design could also cause distortion of the point spread function (PSF), especially near the edges. Figure \ref{fig:psf} shows the spatial distribution of the average PSF, measured in a $500\times500~\mathrm{pixel^2}$ grid on a 180~s calibrated exposure. In each grid, we selected up to 20 bright, non-saturated stars to construct an average PSF. The system's large pixel scale ($2^{\prime\prime}.97~\mathrm{pixel^{-1}}$) results in an undersampled PSF; consequently, $>99\ \%$ of the source signal falls within a $3\times3~\mathrm{pixel^2}$ ($9\times9~ \mathrm{arcsec}^2$) area, largely independent of seeing conditions. We fitted the PSF with 1D Gaussian functions along the x and y directions after resampling it onto a $2\times2$ sub-pixel grid. The resulting PSF characteristics are consistent across the FoV, achieving a FWHM $<2$ pixels (6 arcsec) and an ellipticity $b/a\sim0.9$ in all grids. Given the relatively large pixel scale of SOLO, aperture photometry is performed using a large aperture radius (typically $\approx 4$ pixels). As a result, any moderate degradation of the PSF toward the field edges in the RASA-11 optical system has a negligible impact on the photometric measurements. We conclude that SOLO maintains a consistent optical performance across its field of view for a 180~s exposure.

\subsection{Photometric calibration}\label{subsec:calibration}
\begin{figure}
\centering
\includegraphics[width=0.95\linewidth]{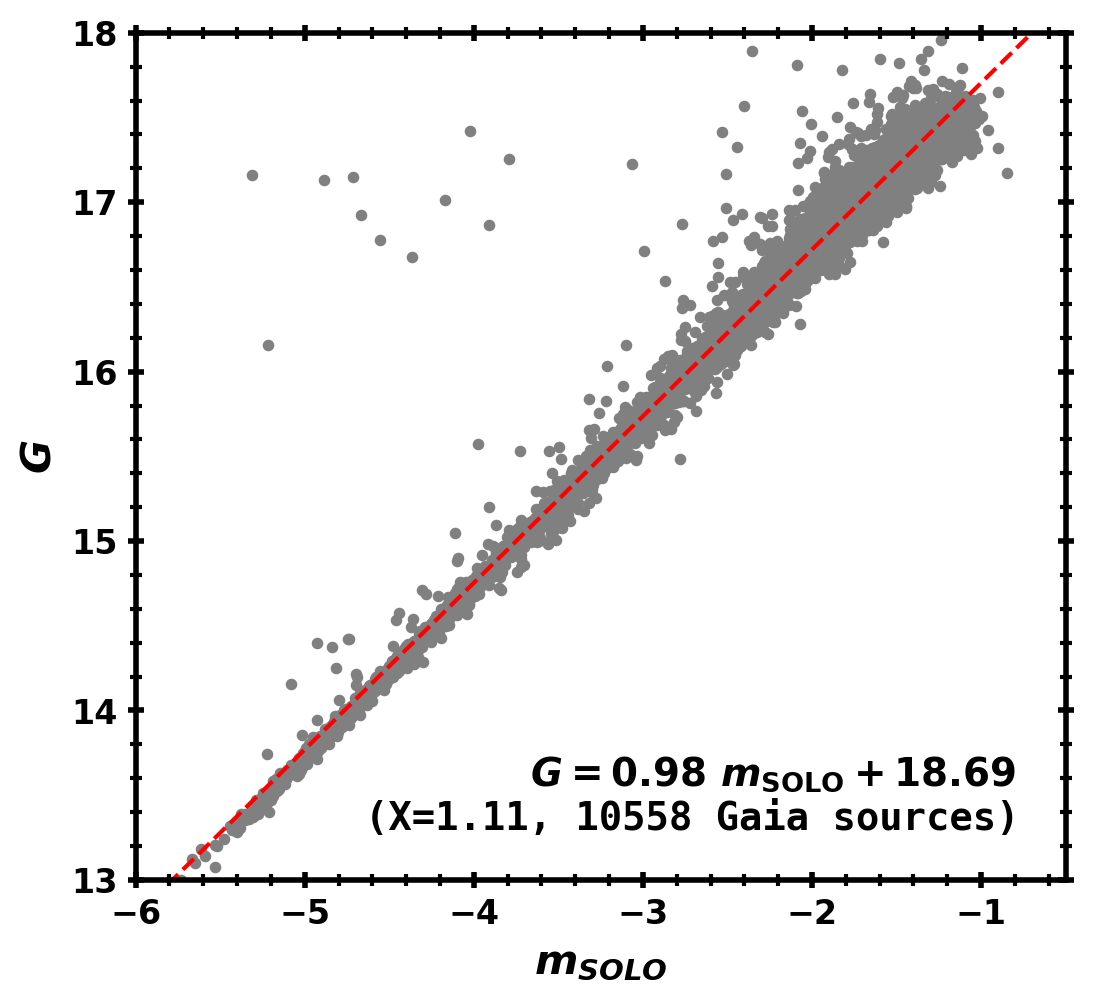}
\caption{Results of aperture photometry for 10,062 \textit{Gaia} sources ($13<G<18$) within the SOLO field of view, based on data obtained on July 22, 2025, at an airmass of $X=1.1$. The red dotted line represents a linear fit to the data. The instrumental magnitudes ($m_\mathrm{SOLO}$) demonstrate a strong correlation with the \textit{Gaia} $G$ magnitudes, with a fitted slope near unity. The estimated photometric zero point for this field is $z \approx 18.7$ mag. \label{fig:zp}}
\end{figure}

\begin{figure}
\centering
\includegraphics[width=\linewidth]{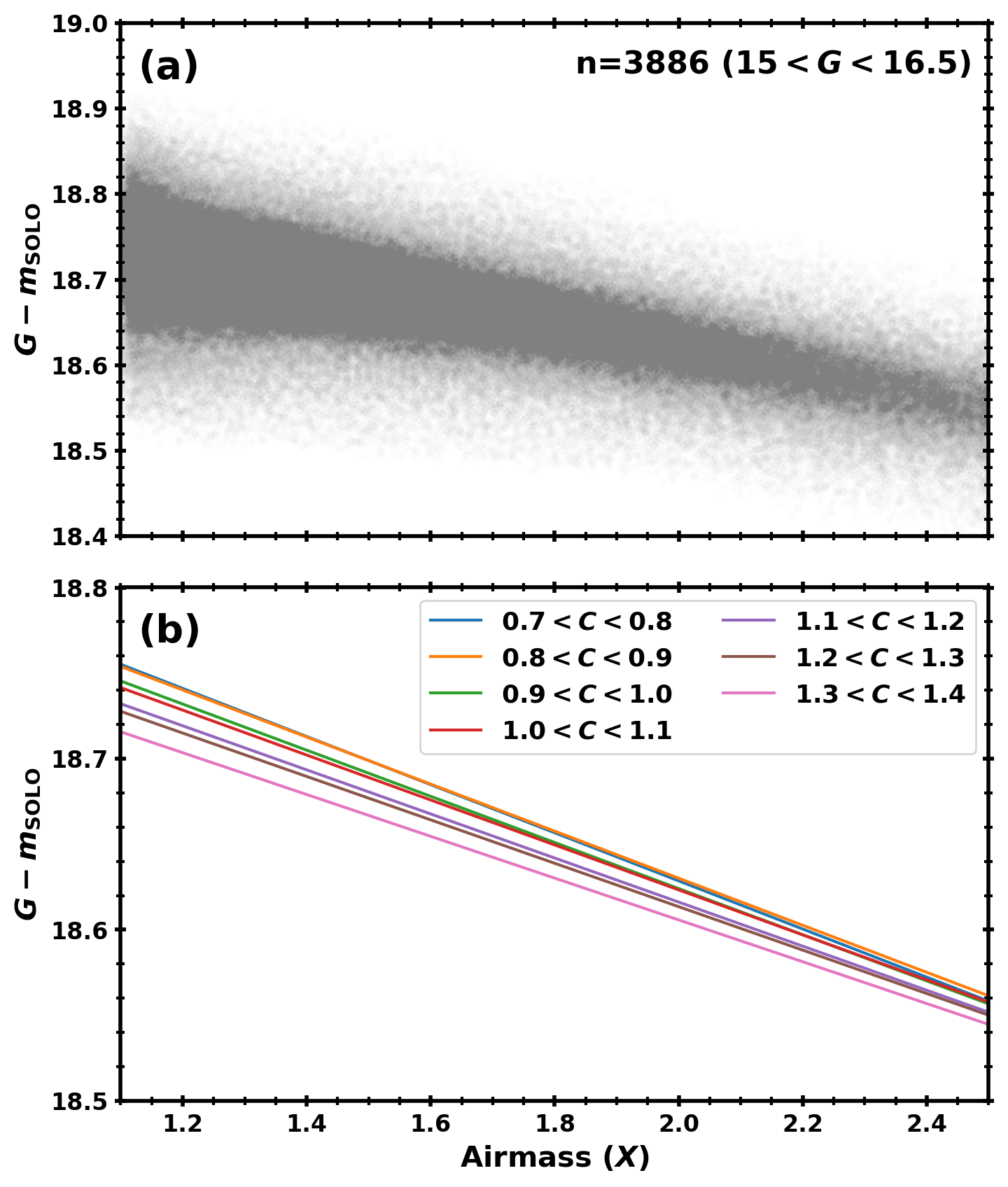}
\caption{Dependence of the photometric zero point on airmass ($X$) and color ($C=G_{BP}-G_{RP}$), using data obtained on July 22, 2025. (a) Airmass dependence of the zero point ($G-m_\mathrm{SOLO}$) for 3,886 \textit{Gaia} sources within the magnitude range $15 < G < 16.5$. (b) Same as panel (a), but with the sources divided into seven color bins ($0.7 < C < 1.4$). The solid lines represent linear fits to each bin, illustrating that the zero point depends on both the airmass and the stellar colors. \label{fig:extinction}}
\end{figure}

\begin{figure}
\centering
\includegraphics[width=0.95\linewidth]{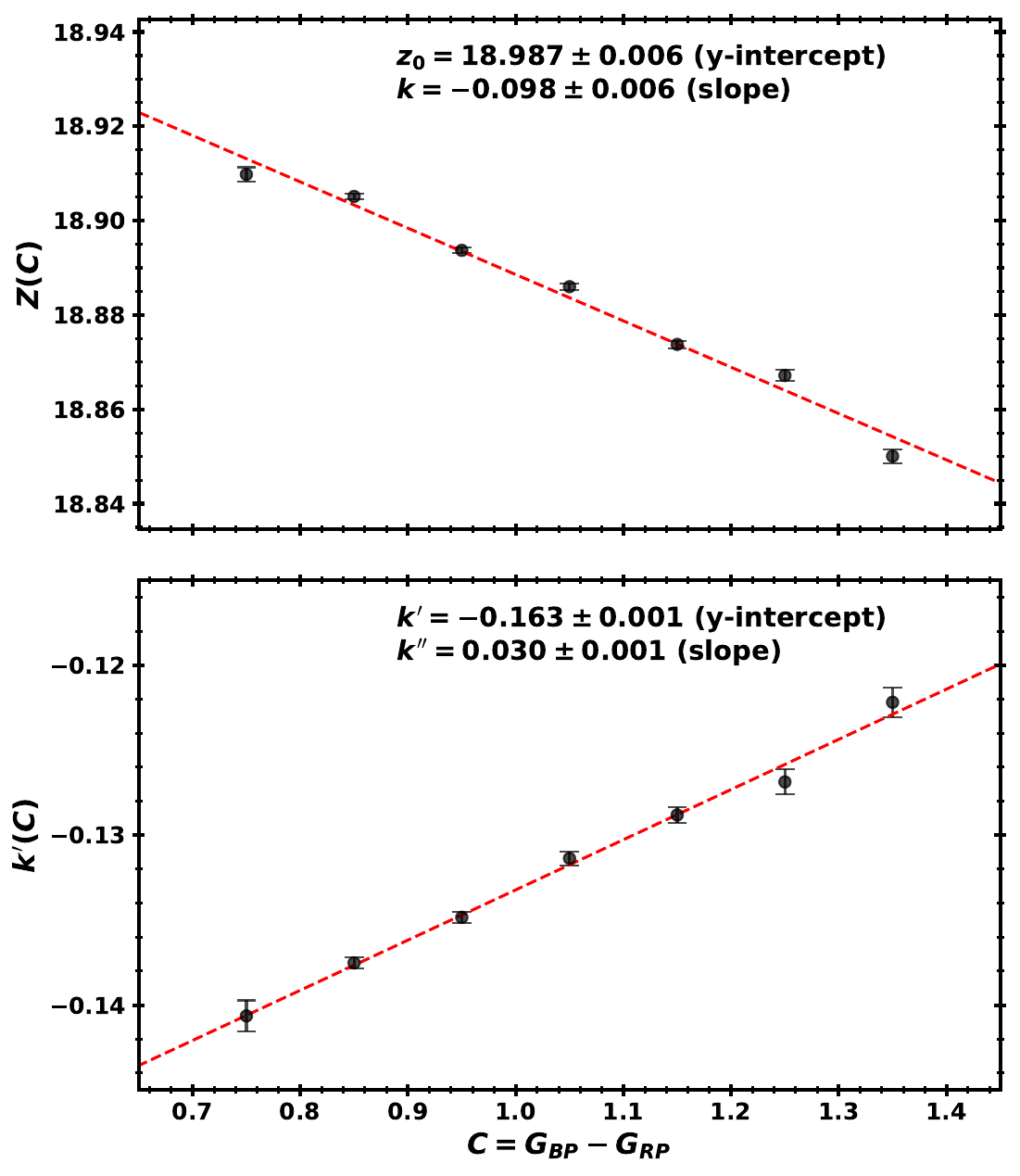}
\caption{Stellar color dependency of the derived atmospheric coefficients ($z_0$, $k$, $k^\prime$, and $k^{\prime\prime}$). The upper panel shows the zero-airmass intercept ($z_0$) as a function of the color ($C = G_{BP}-G_{RP}$), where the fitted slope corresponds to the color term ($k$). The lower panel shows the first-order atmospheric coefficient ($k^\prime$) as a function of the color, where the fitted slope corresponds to the second-order extinction coefficient ($k^{\prime\prime}$).\label{fig:color}}
\end{figure}

\begin{figure*}
\centering
\includegraphics[width=\linewidth]{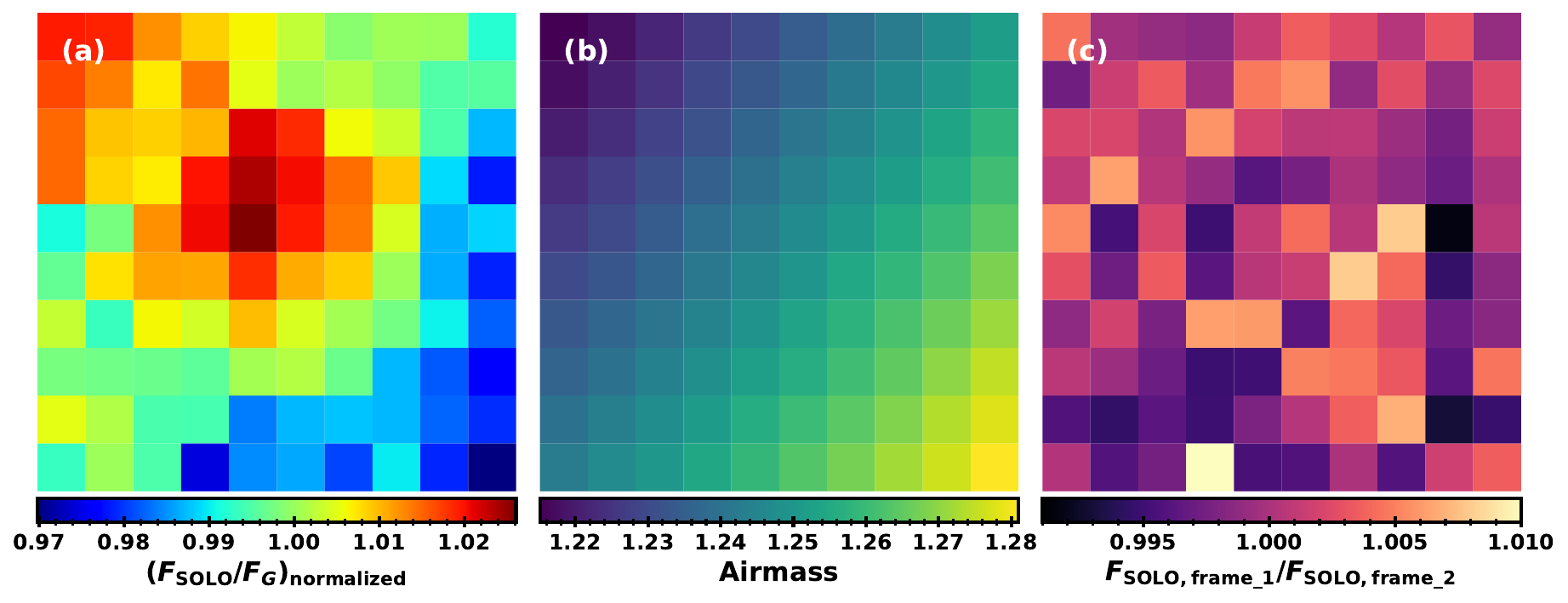}
\caption{Spatial distribution of photometric zero points across the SOLO field of view, derived from 9,929 field stars taken on July 22, 2025 (airmass $X \approx 1.25$, 180~s exposure). (a) Normalized flux throughput. Zero points were median-estimated within $400 \times 400$ pixel subregions and subsequently converted to relative flux ratios using the relation $F_\mathrm{SOLO}/F_G = 10^{-0.4(m_\mathrm{SOLO}-G)}$. (b) Variation in airmass across the field of view. (c) Flux ratio between identical field stars measured in two consecutive SOLO frames to demonstrate short-term photometric stability. \label{fig:fieldzp}}
\end{figure*}

As described in Section~\ref{subsec:telescope_and_detector}, SOLO uses a clear filter to maximize throughput, providing a wide sensitive wavelength range ($\lambda \approx$ 300--900 nm).
Considering that atmospheric scattering reduces transmission at shorter wavelengths, the effective spectral response of SOLO is expected to be roughly comparable to that of the \textit{Gaia} photometric $G$-band ($\lambda \approx 400$--$900$ nm).
In this section, we convert the instrumental magnitudes obtained by SOLO into \textit{Gaia} $G$ magnitudes by accounting for stellar colors, and evaluate how accurately the resulting calibrated magnitudes agree with the \textit{Gaia} $G$-band photometric system, which serves as the standard reference.

The photometric zero point of SOLO was estimated using the \textit{Gaia} system through the following transformation:
\begin{equation} \label{eq:photparams}
G-m_\mathrm{SOLO} = z_0+k'X+(k+k''X)C,~
\end{equation}
where $G$ is the \textit{Gaia G}-magnitude of the field stars, $m_\mathrm{SOLO}$ is their instrumental magnitudes ($-2.5\log_{10}{(\mathrm{ADU})}$), $z_0$ is the photometric zero point at airmass $X=0$ and color $C=0$, $k'$ and $k''$ are the first and second-order atmospheric extinction coefficients, respectively, and $k$ is the color term at $X=0$. We adopted the $C:=G_{BP}-G_{RP}$, where $G_{BP}$ and $G_{RP}$ are the \textit{Gaia} blue and red bands, respectively \citep{2023A&A...674A...1G}.

To adapt Equation (\ref{eq:photparams}) to our system, we first examined the suitability of the \textit{Gaia} system by checking the consistency between the instrumental magnitude ($m_\mathrm{SOLO}$) and the \textit{Gaia} $G$ magnitude in a single 180~s exposure (taken on July 22, 2025, at an elevation of 64\degree, $X=1.11$). We performed aperture photometry with a 4-pixel radius (approximately twice the FWHM) on the field stars, selecting those that: i) are the \textit{Gaia} sources ii) contained no bad pixels in the aperture, iii) were not saturated ($G>13$ mag), and iv) were isolated (no other \textit{Gaia} sources found within $30^{\prime\prime}$). This selection yielded 10,558 sources in the field. We find that $m_\mathrm{SOLO}$ versus $G$ magnitudes shows a linear correlation over the analyzed magnitude range ($13 < G < 18$), with a slope near unity and a zero point ($z=G-m_\mathrm{SOLO}$) of $z \approx 18.7$ mag at $X=1.11$ (Figure~\ref{fig:zp}).

Next, we determined the atmospheric coefficients ($k, k^\prime$, and $k^{\prime\prime}$) by observing a single field over a wide airmass range ($X=1.1$--$2.5$). In Equation (\ref{eq:photparams}), the atmospheric coefficients ($k^\prime$ and $k^{\prime\prime}$) vary with time depending on atmospheric conditions, whereas $z_0$ and $k$ primarily reflect instrument-specific properties. To accurately determine these instrument-related parameters, we first constrain the airmass-dependent terms by analyzing data obtained over a wide airmass range. We analyzed the zero point for 3,886 \textit{Gaia} sources with $15<G<16.5$ in this field. As shown in Figure \ref{fig:extinction}a, the zero point demonstrates a general dependence on airmass. The sources were then divided into stellar color bins from $C=0.7$ to $C=1.4$ (0.1 mag bin size), where $C := G_{BP}-G_{RP}$, with each bin containing 106 to 954 sources. The results are shown in Figure \ref{fig:extinction}b. It is found that the photometric zero point ($G-m_\mathrm{SOLO}$) clearly depends on both airmass (the $k^\prime$ term) and color (the $k$ and $k^{\prime\prime}$ terms). In Figure \ref{fig:extinction}b, $k$ represents the color dependency of the y-intercept, while $k^{\prime\prime}$ represents the color dependency of the slope. The dependence of the zero-airmass intercept ($z_0$) and $k^\prime$ on color $C$ is shown in Figure~\ref{fig:color}, where we explicitly indicate these parameters as a function of $C$.

From this dataset, we derived the parameters in Equation (\ref{eq:photparams}) as: $z_0=18.987\pm0.006$, $k=-0.098\pm0.006$, $k'=-0.163\pm0.001$, and $k''=0.030\pm0.001$. These results demonstrate that to achieve photometric accuracy better than 0.1 mag, variations in airmass and color must be taken into account. In practice, our data calibration pipeline (Section \ref{subsec:operation}) accomplishes this by performing differential photometry. For each asteroid, it selects nearby \textit{Gaia} sources with similar colors ($0.9<C<1.1$) to define a local zero point and calibrate the asteroid's magnitude.

Additionally, we noticed small spatial variations in the photometric zero point across the detector. Figure \ref{fig:fieldzp} illustrates these variations using 9,929 field stars observed on July 22, 2025, with values median-estimated within $400 \times 400$ pixel bins. Figure \ref{fig:fieldzp}a indicates that residual differences between the \textit{Gaia} $G$ magnitudes and the calibrated SOLO magnitudes remain $\lesssim 3$\% across the field of view, including regions beyond the nominal image circle of the RASA-11. This residual variation is likely a consequence of the large FoV ($3.4^\circ \times 3.4^\circ$), where airmass gradients primarily drive throughput variations. The maximum airmass difference across the field is $\Delta X \approx 0.06$ (Figure \ref{fig:fieldzp}b), which corresponds to a flux variation of approximately $\lvert{k^\prime \Delta X \rvert}\approx 0.01$ mag (or $\sim 1\%$) assuming an extinction coefficient of $k^\prime = 0.16$. Localized atmospheric variations, such as thin cirrus, may introduce additional zero-point fluctuations of $\sim 1\%$ with a random spatial distribution (Figure \ref{fig:fieldzp}c). Our analysis demonstrates that under stable weather conditions, a photometric accuracy better than approximately 3\% can be achieved across the entire field of view.

\subsection{Limiting magnitude}\label{subsec:limitmag}
\begin{figure*}
\centering
\includegraphics[width=\linewidth]{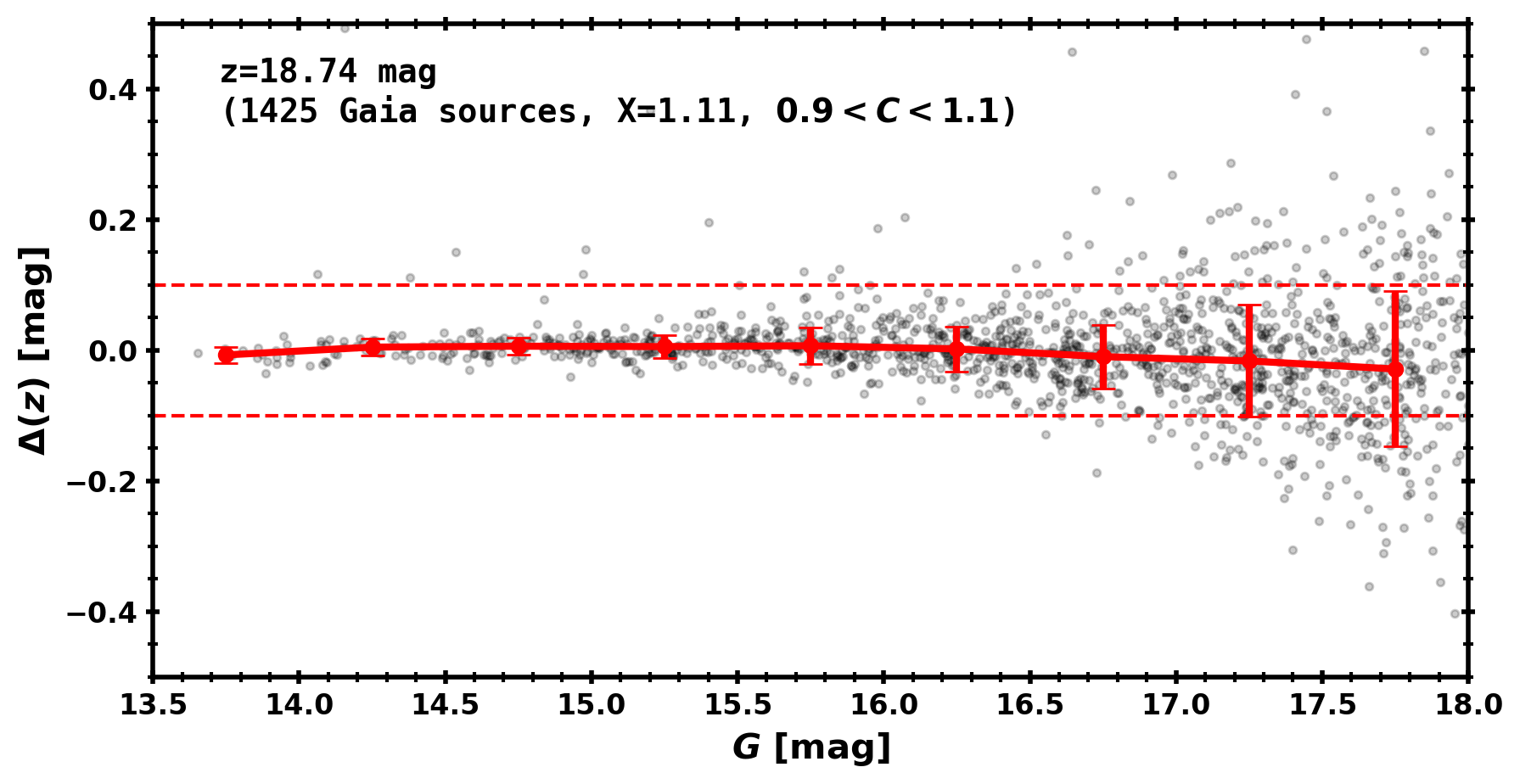}
\caption{Deviation of field star magnitudes from the derived photometric zero point as a function of \textit{Gaia} $G$-magnitude, using data obtained on July 22, 2025 (airmass $X=1.11$; altitude 64\degree). Black points represent 1,425 field stars within the central $1500 \times 1500$ pixel region that satisfy the color constraint $0.9 < G_{BP} - G_{RP} < 1.1$. The red solid line and error bars indicate the median zero-point trend and its $1\sigma$ scatter, calculated in 0.5 mag intervals. The red dotted lines mark the $\pm0.1$ mag deviation limits relative to the central zero point. \label{fig:limitmag}}
\end{figure*}

In this subsection, we investigate the practical limiting magnitude of SOLO under real operational conditions. The SOLO operates with open-loop tracking, without an auto-guiding system.
As a result, long integrations ($\gtrsim 10$~min) are not feasible due to tracking errors arising from the worm-gear accuracy and other mechanical instabilities, such as the mount and tripod.
Conversely, very short exposures are also not desirable when considering data volume and readout noise. We therefore adopt a representative integration time of 180~s, which lies within the realistic operational range (60--300~s), to evaluate the limiting magnitude of SOLO.

Here, we define the limiting magnitude as the brightness at which a signal-to-noise ratio (S/N) of 10 is achieved for a 180~s exposure. This corresponds to a photometric error of $10\%$ of the signal, or $\sigma\approx 0.1$ mag. 
This threshold is motivated by two considerations.
First, it is sufficient to detect typical rotational brightness variations of asteroids, whose median light curve amplitudes are of 0.2--0.3 mag \citep{2009Icar..202..134W}. 
Second, {\spherex} targets an overall calibration accuracy of $\sim 3\%$ \citep{2025arXiv251102985B}. Since there are additional complications specific to moving objects, which are not point sources, the SSOC aims for an inter-band spectral calibration accuracy at the $\sim 10\%$ level for $\sim 19$ ABmag objects, making this level of photometric precision a natural requirement for ancillary ground-based observations.
We note that the necessity for light curve correction diminishes for asteroids with amplitudes significantly below this threshold.

To empirically determine the limiting magnitude, we measured the deviation of field star instrumental magnitudes from the derived zero points as a function of their \textit{Gaia} $G$ magnitudes. Figure \ref{fig:limitmag} displays a representative plot of these deviations for a single exposure, restricted to stars within the central $1500 \times 1500$ pixel region and the color range $0.9 < C < 1.1$. Our results show that SOLO typically reaches a limiting magnitude of $G \approx 17$--$18$ mag at the $\sigma = 0.1$ mag level. The exact performance varies with sky conditions and the star's position on the sensor, with lower sensitivity at the edges due to vignetting. The best performance measured during commissioning was $G = 17.8 \pm 0.2$ mag.

\subsection{Sample light curves}\label{subsec:sample_lightcurve}

\begin{figure*}[!t]
\centering
\includegraphics[width=\linewidth]{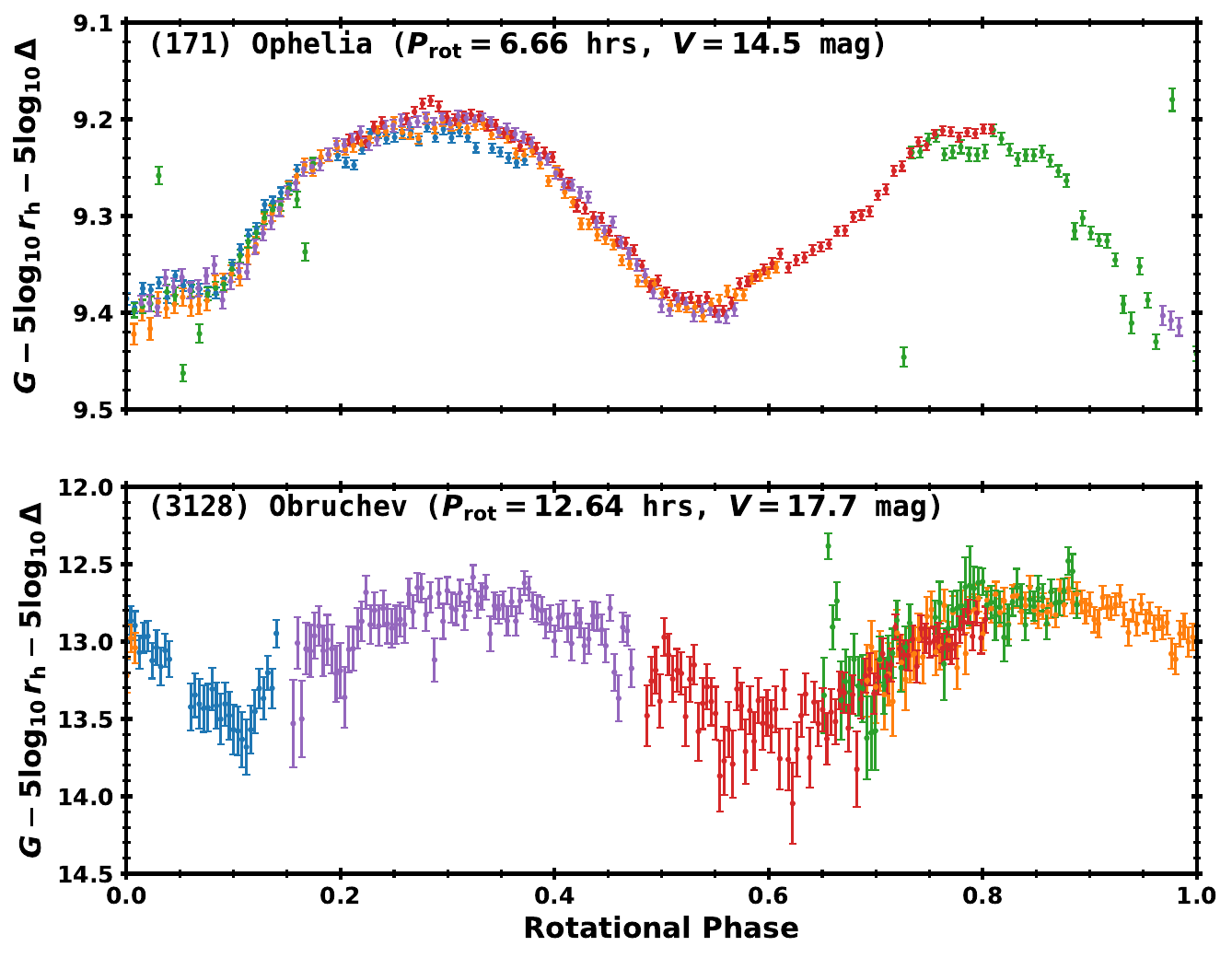}
\caption{Sample light curves of two asteroids, 171 Ophelia (top) and 3128 Obruchev (bottom), as a result of the absolute photometry. The blue, orange, green, red, and purple colors correspond to the data points on August 23, 25, 26, 27, and 30, respectively. \label{fig:lightcurves}}
\end{figure*}

To test whether the photometric accuracy evaluated in Section~\ref{subsec:limitmag} can be reliably applied to asteroid light curve observations, we conducted continuous monitoring of the same sky region near the ecliptic plane over about a one-week period during consecutive clear nights. We conducted sample light curve observations during the new-moon phase between August~23 and 30. During this period, we targeted a single region centered at $\mathrm{RA}=\mathrm{02^h\,42^m\,30^s}$ and $\mathrm{Dec}=\mathrm{12^\circ\,44'\,24''}$ (ecliptic coordinates $l=42^\circ.16$, $b=-2^\circ.88$). Approximately 15 asteroids with $V<18$~mag were detected in SOLO's FoV.

For these sample asteroids, we performed time-series absolute photometry using the data reduction pipeline described in Sections~\ref{subsec:operation} and~\ref{subsec:calibration}. To correct for viewing-geometry effects, the magnitudes obtained on different nights were adjusted for the heliocentric distance ($5\log (r_\mathrm{h}/1~\mathrm{au})$) and the geocentric distance ($5\log (\Delta/1~\mathrm{au})$). For these targets, we did not apply a phase-angle correction, as the variation in phase angle during the observation period was negligible ($\Delta \alpha= 0.2-1.0\degree$ during this period).

Figure~\ref{fig:lightcurves} shows the resulting light curves for two representative asteroids
(one bright target with sufficiently high $S/N$ and one faint target close to the limiting magnitude): (171)~Ophelia ($P_\mathrm{rot}=6.7$~hours\footnote{Rotation period retrieved from the DAMIT database: \url{https://damit.cuni.cz/}} and $V=14.5$~mag\footnote{Visual magnitude retrieved from JPL/Horizons: \url{https://ssd.jpl.nasa.gov/horizons/}})
and
(3128)~Obruchev ($P_\mathrm{rot}=12.7$~hours and $V=17.7$~mag).

For both objects, the light curves obtained on different nights overlap remarkably well at the same rotational phase, despite being calibrated independently on each night.
This demonstrates that SOLO provides consistent absolute photometry across multiple nights.
For the brighter asteroid (171)~Ophelia, which has a S/N of $\sim100$, the light curves agree at the $\sim1\%$ level, consistent with the photometric precision estimated in Section~\ref{subsec:limitmag}.

For the fainter asteroid (3128)~Obruchev ($V=17.7$~mag), a scatter of order $\sim0.1$~mag is observed, as expected for $S/N\sim10$, while the median magnitudes remain consistent between nights. For this asteroid, we further examined the mean magnitude within a restricted rotational phase range (0.7--0.8 in Figure~\ref{fig:lightcurves}) across three different nights. We found that the night-to-night scatter of the averaged magnitudes is $\lesssim$~0.03~mag, significantly smaller than the single-measurement uncertainty ($\sim0.1$~mag), and consistent with the expected reduction of random errors by $1/\sqrt{N}$ ($\sim0.1/\sqrt{N}\sim0.03$, where $N\sim10$ is the number of samples per night). This indicates that systematic calibration errors are effectively suppressed.

These empirical results demonstrate that systematic effects are properly corrected by the calibration pipeline and validate that the limiting-magnitude estimate derived in Section~\ref{subsec:limitmag} is appropriate for practical asteroid light curve observations with SOLO.

\section{SOLO operations toward the {\spherex} SSOC}
\label{sec:science_programs}

In this section, we describe the operational concept of SOLO in support of the {\spherex} Solar System Object Catalog (SSOC) and discuss how ground-based light curve observations are essential for maximizing the scientific value of {\spherex} asteroid spectra.

Launched in March 2025, {\spherex} is expected to obtain near-infrared ($\lambda = 0.7$--$5.0~\mu$m) spectra for approximately $\sim10^{5}$ asteroids \citep{2022Icar..37114696I}. 
This unprecedented dataset enables statistical studies of asteroid surface compositions, including hydrated minerals and silicates. However, asteroid brightness variations due to rotation typically reach the level of several to tens of percent, and therefore accurate correction for rotational light curve effects is required in order to derive reliable reflectance spectra.

At present, only a limited number of asteroids have well-constrained rotational states and shape models. For example, the DAMIT database contains only $\sim \mathrm{300}$ asteroids with reliable light curve shape models above a given quality threshold (\texttt{quality flag} $\geq2.5$, as of January 2026). Consequently, despite the large number of asteroid spectra expected from \spherex, a substantial fraction of the SSOC data would remain difficult to interpret quantitatively without additional light curve information. SOLO is designed to address this limitation by providing absolutely calibrated optical light curves through multi-night, continuous monitoring of selected sky regions. In particular, SOLO can deliver the time-series brightness (i.e., light curve) on a timescale of approximately 1–2 weeks
for asteroids observed by \spherex. It nominally takes 2 weeks for {\spherex} to obtain the full spectrum of a fixed sky region, and the light curves of that timescale are necessary. For asteroids with existing DAMIT shape models, SOLO light curves can further be used to correct phase offsets caused by small uncertainties in the rotational period, thereby improving the accuracy of rotational phase assignment at the {\spherex} observation epochs.

The observing strategy of SOLO for SSOC support is still under refinement and will be optimized based on accumulated operational data. Parameters such as exposure time, cadence, and survey area will be tuned to balance photometric precision and sky coverage. As a baseline concept, SOLO will repeatedly observe selected sky regions that overlap with {\spherex} pointings, targeting periods before and after the {\spherex} observations and monitoring the same field continuously for approximately one week. With a representative exposure time of 180~sec, SOLO is expected to detect roughly $\sim15$ asteroids per field down to $G \sim 18$ mag near the ecliptic plane. Since {\spherex} observes both the leading and trailing sides relative to the Sun, this strategy may enable the acquisition of light curves for up to $\sim1500$ asteroids per year, although the actual number will depend on the final observing configuration.

The scientific impact of this synergy can be illustrated by comparison with previous surveys. The AKARI mission provided reflectance spectra beyond 2.5~$\mu$m for 66 asteroids \citep[]{2019PASJ...71....1U}, which remains the largest homogeneous dataset in this wavelength range to date. By enabling reliable rotational corrections for a substantially larger number of {\spherex} asteroid spectra, SOLO has the potential to increase the number of quantitatively usable reflectance spectra beyond 2.5~$\mu$m by more than an order of magnitude.

The {\spherex} SSOC is an officially coordinated activity within the {\spherex} mission framework, supported by a formal understanding with NASA. A comprehensive description of the SSOC catalog definition, construction, and scientific scope will be presented in a dedicated future publication.


\acknowledgments

This research was supported by a National Research Foundation of Korea (NRF) grant funded by the Korean government (MEST; No. 2023R1A2C1006180). Bumhoo Lim was partially supported for the research activity at Seoul National University by the Basic Science Research Program through the NRF, funded by the Ministry of Education (RS-2025-25436385). We are grateful to Dr. Ko Arimatsu (National Astronomical Observatory of Japan) for helpful advice and discussions on the initial installation of the RASA-11 system.






\end{document}